%%%%%%%%%%%%%%%%%%%%%%%%%%%%%%%%%%%%%%%%%%%%%%%%%%%%%%%%%%%%%%%%%%%%
%  TeX Definitions                                                 %
%%%%%%%%%%%%%%%%%%%%%%%%%%%%%%%%%%%%%%%%%%%%%%%%%%%%%%%%%%%%%%%%%%%%

\newif\iffigs\figstrue
% Uncomment the next line if you do not want the figures
%\figsfalse

%
% the following is to use blackboard bold fonts --
\let\useblackboard=\iftrue
%
% activate this if you don't have them.
%\let\useblackboard=\iffalse
%
% You might also need to remove this line.
\newfam\black

\input harvmac.tex

\iffigs
  \input epsf
\else
  \message{No figures will be included.  See TeX file for more
information.}
\fi

\def\Title#1#2{\rightline{#1}
\ifx\answ\bigans\nopagenumbers\pageno0\vskip1in%
\baselineskip 15pt plus 1pt minus 1pt
\else%\special{papersize=11in,8.5in}%
\def\listrefs{\footatend\vskip 1in\immediate\closeout\rfile\writestoppt
\baselineskip=14pt\centerline{{\bf References}}\bigskip{\frenchspacing%
\parindent=20pt\escapechar=` \input
refs.tmp\vfill\eject}\nonfrenchspacing}
\pageno1\vskip.8in\fi \centerline{\titlefont #2}\vskip .5in}

\ifx\answ\bigans\def\tcbreak#1{}\else\def\tcbreak#1{\cr&{#1}}\fi
\useblackboard
\message{If you do not have msbm (blackboard bold) fonts,}
\message{change the option at the top of the tex file.}
\font\blackboard=msbm10 %scaled \magstep1
\font\blackboards=msbm7
\font\blackboardss=msbm5
%\newfam\black
\textfont\black=\blackboard
\scriptfont\black=\blackboards
\scriptscriptfont\black=\blackboardss
\def\Bbb#1{{\fam\black\relax#1}}
\else
\def\Bbb#1{{\bf #1}}
\fi
% *************************************
%
\def\yboxit#1#2{\vbox{\hrule height #1 \hbox{\vrule width #1
\vbox{#2}\vrule width #1 }\hrule height #1 }}
\def\fillbox#1{\hbox to #1{\vbox to #1{\vfil}\hfil}}
\def\ybox{{\lower 1.3pt \yboxit{0.4pt}{\fillbox{8pt}}\hskip-0.2pt}}
\def\np#1#2#3{Nucl. Phys. {\bf B#1} (#2) #3}

\def\mpl#1#2#3{Mod. Phys. Lett. {\bf #1} (#2) #3}

\def\comments#1{}

\def\half{{1\over 2}}

\def\CP{{\cal P}}

\def\a{\alpha}

\def\II{\relax{I\kern-.07em I}}

\def\IZ{\relax\ifmmode\mathchoice
{\hbox{\cmss Z\kern-.4em Z}}{\hbox{\cmss Z\kern-.4em Z}}
{\lower.9pt\hbox{\cmsss Z\kern-.4em Z}}
{\lower1.2pt\hbox{\cmsss Z\kern-.4em Z}}\else{\cmss Z\kern-.4em
Z}\fi}
\def\IB{\relax{\rm I\kern-.18em B}}

\def\ID{\relax{\rm I\kern-.18em D}}
\def\IE{\relax{\rm I\kern-.18em E}}
\def\IF{\relax{\rm I\kern-.18em F}}
\def\IG{\relax\hbox{$\inbar\kern-.3em{\rm G}$}}
\def\IGa{\relax\hbox{${\rm I}\kern-.18em\Gamma$}}
\def\IH{\relax{\rm I\kern-.18em H}}
\def\II{\relax{\rm I\kern-.18em I}}
\def\IK{\relax{\rm I\kern-.18em K}}
\def\IP{\relax{\rm I\kern-.18em P}}
%\def\IX{\relax{\rm X\kern-.01em X}}
%this doesn't work

\useblackboard
\def\IZ{\relax\Bbb{Z}}
\fi

\font\cmss=cmss10 \font\cmsss=cmss10 at 7pt
\def\IR{\relax{\rm I\kern-.18em R}}

\def\BR{\IR}
\def\BZ{\IZ}
\def\BR{\IR}

%%%%%%%%%%%%%%%%%%%%%%%%%%%%%%%%%%%%%%%%%%%%%%%%%%%%%%%%%%%%%%%%%%%%%%%%%%%%
%                    F I G U R E S                                         %
%%%%%%%%%%%%%%%%%%%%%%%%%%%%%%%%%%%%%%%%%%%%%%%%%%%%%%%%%%%%%%%%%%%%%%%%%%%%
%\figsfalse

\iffigs
  \input epsf
\else
  \message{No figures will be included.  See TeX file for more
information.}
\fi

%%% \iffigs
%%% \midinsert
%%% $$\vbox{\centerline{\epsfxsize=4in\epsfbox{figa.eps}}
%%% \centerline{Figure 1. $E_2$ and related theories.}}$$
%%% \endinsert
%%% \fi

%%%%%%%%%%%%%%%%%%%%%%%%%%%%%%%%%%%%%%%%%%%%%%%%%%%%%%%%%%%%%%%%%%%%%%%%%%%%
%                    Definitions from LaTeX                                %
%%%%%%%%%%%%%%%%%%%%%%%%%%%%%%%%%%%%%%%%%%%%%%%%%%%%%%%%%%%%%%%%%%%%%%%%%%%%

%%%
%%% All those have problems with Font \rm
%%%

\def\log{{{\rm log}\,}}
\def\logx#1{{{\rm log}\,{\left({#1}\right)}}}

\def\lim{{lim}}

%%%%%%%%%%%%%%%%%%%%%%%%%%%%%%%%%%%%%%%%%%%%%%%%%%%%%%%%%%%%%%%%%%%%%%%%%%%%
%                    My definitions                                        %
%%%%%%%%%%%%%%%%%%%%%%%%%%%%%%%%%%%%%%%%%%%%%%%%%%%%%%%%%%%%%%%%%%%%%%%%%%%%
\input epsf

\def\SUSY#1{{{\cal N}= {#1}}}                   % N=? SUSY
\def\lbr{{\lbrack}}                             % [
\def\rbr{{\rbrack}}                             % ]

\def\wdg{{\wedge}}                              % wedge product

                              % Wilson lines

\def\inv#1{{1\over{#1}}}                              % inverse
                           % O(x)

\def\MR#1{{{\BR}^{#1}}}               % Real numbers
               % Complex numbers

%%% \def\MR#1{{{\bf R}^{#1}}}               % Real numbers
%%% \def\MC#1{{{\bf C}^{#1}}}               % Complex numbers
\def\MR#1{{{\BR}^{#1}}}               % Real numbers
               % Complex numbers
\def\MS#1{{{\bf S}^{#1}}}               % Circle, sphere,...
               % disk, ball,...
\def\MT#1{{{\bf T}^{#1}}}               % Torus
\def\CP#1{{{\bf P}^{#1}}}              % CP
\def\MF#1{{{\bf F}_{#1}}}               % Ruled surface F_n

             % Patch
                    % line-bundle
              % derivative

                 % Left large bracket
                % Right large bracket
              % SL(*,Z)

                             % identity matrix

      % commutator
               % anti-commutator

\def\ev#1{{\langle {#1} \rangle}}           % expectation value
    % expectation value of trace

      % trace
\def\trp#1{{{\rm tr}\{ {#1} \} }}            % trace
            % Trace
\def\trr#1#2{{{\rm tr}_{#1}\{ {#2} \} }}            % trace in a rep
\def\Trr#1#2{{{\rm Tr}_{#1}\{ {#2} \} }}            % Trace in a rep

\def\rep#1{{{\bf {#1}}}}                      % representation
                  % Imaginary
                  % Imaginary

%\def\widebar#1{{\bar{#1}}}                    % Wide bar
\def\widebar#1{{\overline{#1}}}                    % Wide bar
                 % Pauli matrix

\def\Ol#1{{ {\cal O}({#1}) }}                      % correction O()
                     % Normal bundle

                      % Hodge star
                         % sign
\def\hepth#1{{\it hep-th/{#1}}}

\def\frac#1#2{{{{#1}}\over {{#2}}}}           % {} over {}

%%%%%%%%%%%%%%%%%%%%%%%%%%%%%%%%%%%%%%%%%%%%%%%%%%%%%%%%%%%%%%%%%%%%%%%%%%%%
%                    Greek                                                 %
%%%%%%%%%%%%%%%%%%%%%%%%%%%%%%%%%%%%%%%%%%%%%%%%%%%%%%%%%%%%%%%%%%%%%%%%%%%%
\def\u{{\mu}}
\def\v{{\nu}}

\def\lam{{\lambda}}

\def\Lam{{\Lambda}}

%%% \def\Dsh{{D\!\!\!\slash}}     % D slash

%%%%%%%%%%%%%%%%%%%%%%%%%%%%%%%%%%%%%%%%%%%%%%%%%%%%%%%%%%%%%%%%%%%%%%%%%%%%
%     Special Purpose  Definitions                                         %
%%%%%%%%%%%%%%%%%%%%%%%%%%%%%%%%%%%%%%%%%%%%%%%%%%%%%%%%%%%%%%%%%%%%%%%%%%%%
                            % 2\times 2  J

\def\wu{{\widebar{u}}}

                                % Wave function propagator

\def\wbu{{\widebar{u}}}
\def\wbx{{\widebar{x}}}
\def\wby{{\widebar{y}}}

%%% ------------------------- CUT HERE ---------------------------------%

%%%%%%%%%%%%%%%%%%%%%%%%%%%%%%%%%%%%%%%%%%%%%%%%%%%%%%%%%%%%%%%%%%%%%%%%%%%%
%                    TITLE PAGE                                            %
%%%%%%%%%%%%%%%%%%%%%%%%%%%%%%%%%%%%%%%%%%%%%%%%%%%%%%%%%%%%%%%%%%%%%%%%%%%%

%%% \draftmode

%
\Title{ \vbox{\baselineskip12pt\hbox{hep-th/9710053, PUPT-1728}}}
{\vbox{
\centerline{Correlators}
\centerline{of the Global Symmetry Currents}
\centerline{of 4D and 6D Superconformal Theories}}}
\centerline{Yeuk-Kwan E. Cheung$^{1}$,
           Ori J. Ganor$^{2}$ and Morten Krogh$^{3}$}
\smallskip
\smallskip
\centerline{Department of Physics, Jadwin Hall}
\centerline{Princeton University}
\centerline{Princeton, NJ 08544, USA}
\centerline{\tt ${}^1$cheung@viper.princeton.edu}
\centerline{\tt ${}^2$origa@puhep1.princeton.edu}
\centerline{\tt ${}^3$krogh@phoenix.princeton.edu}
%%%
\bigskip
\bigskip
\noindent
We study the two-point correlators of the currents of the $E_8$
global symmetry in the $\SUSY{(1,0)}$ superconformal six-dimensional
theory as well as in the 4D superconformal theories
%%% obtained from it by compactification.
upon toroidal compactification.
 From the high-energy behavior of the correlator we deduce that
in 4D 10 copies of the superconformal theory with $E_8$ global symmetry
can be coupled to an $\SUSY{2}$ $E_8$ gauge theory.
 We present three alternative derivations for the expression for
the correlators. One from field-theory, one from M-theory and
one from F-theory.

\Date{October, 1997}

%%% ------------------------- CUT HERE ---------------------------------%

%%%%%%%%%%%%%%%%%%%%%%%%%%%%%%%%%%%%%%%%%%%%%%%%%%%%%%%%%%%%%%%%%%%%
%  B I B L I O G R A P H Y                                         %
%%%%%%%%%%%%%%%%%%%%%%%%%%%%%%%%%%%%%%%%%%%%%%%%%%%%%%%%%%%%%%%%%%%%
\lref\CHS{ C. G. Callan, J. A. Harvey, and A. Strominger,
  {\it ``Supersymmetric String Solitons,''}
  \hepth{9112039}}

\lref\WitCOM{ E. Witten,
  {\it ``Some Comments on String Dynamics,''}
  contributed to Strings '95, \hepth{9507121}}

\lref\StrOPN{A. Strominger,
  {\it ``Open p-Branes,''} \hepth{9512059}}

\lref\WitSML{ E. Witten,
  {``Small Instantons in String Theory,''}
  \np{460}{96}{541}, \hepth{9511030}}

\lref\KKV{S. Katz, A. Klemm and C. Vafa,
  {\it ``Geometric Engineering Of Quantum Field Theories,''}
  \hepth{9609239}}

\lref\BerSad{M. Bershadsky and V. Sadov,
  {\it ``F-Theory on $K3\times K3$ and Instantons on 7-branes,''}
  \hepth{9703194}}

\lref\KMV{A. Klemm, P. Mayr and C. Vafa,
  {\it ``BPS States of Exceptional Non-Critical Strings,''} \hepth{9607139}}

\lref\WitNWT{E. Witten,
  {``Strong Coupling Expansion Of Calabi-Yau Compactification,''}
  \hepth{9602070}}

\lref\WitV{E. Witten,
  {\it ``Physical Interpretation Of Certain Strong Coupling Singularities,''}
  \hepth{9609159}}

\lref\John{J.H. Brodie,
  {\it ``Patterns of Duality in N=1 SUSY Gauge Theories,''} \hepth{9611197}}

\lref\WitBRC{E. Witten,
  {\it ``Solutions Of Four-Dimensional Field Theories Via M Theory,''}
  \hepth{9703166}}

\lref\APSW{ P.C. Argyres, M.R. Plesser,
  N. Seiberg, E. Witten,
  {\it ``New $N=2$ Superconformal Field Theories In Four Dimensions,''}
  \np{461}{96}{71}, \hepth{9511154}}

\lref\MNI{J.A. Minahan and D. Nemeschansky,
  {\it ``An N=2 Superconformal Fixed Point with $E_6$ Global Symmetry,''}
  \hepth{9608047}}

\lref\MNII{J.A. Minahan and D. Nemeschansky,
  {\it ``Superconformal Fixed Points with $E_n$ Global Symmetry,''}
  \hepth{9610076}}

\lref\MNWI{J.A. Minahan, D. Nemeschansky and N.P. Warner,
  {\it ``Investigating the BPS Spectrum of Non-Critical $E_n$ Strings,''}
  \hepth{9705237}}

\lref\MNWII{J. A. Minahan, D. Nemeschansky and N. P. Warner,
  {\it ``Partition Functions for BPS States of the Non-Critical
  $E_8$ String,''}
  \hepth{9707149}}

\lref\MVII{D.R. Morrison and C. Vafa,
  {\it ``Compactifications Of F-Theory On Calabi-Yau Threefolds - II,''}
  \hepth{9603161}}

%%% \lref\HorWit{P. Horava and E. Witten,
%%%   {\it ``Heterotic and Type I String Dynamics from
%%%   Eleven Dimensions,''} preprint IASSNS-HEP-95-86, \hepth{9510209}.}

\lref\HWII{P. Horava and E. Witten,
  {``Eleven-Dimensional Supergravity On a Manifold With Boundary,''}
  \hepth{9603142}}

\lref\GanHan{O.J. Ganor and A. Hanany,
  {``Small $E_8$ Instantons and Tensionless Non-Critical Strings,''}
  \hepth{9602120}}

\lref\GMS{O.J. Ganor, D.R. Morrison, N. Seiberg,
  {\it ``Branes, Calabi-Yau Spaces, and Toroidal Compactification
  of the $N=1$ Six-Dimensional $E_8$ Theory,''} \hepth{9610251}}

\lref\GanTOR{O.J. Ganor,
  {\it ``Toroidal Compactification of Heterotic 6D Non-Critical Strings
  Down to Four Dimensions,''} \hepth{9608109}}

\lref\SWQCD{N. Seiberg and E. Witten,
  {``Monopoles, Duality and Chiral Symmetry Breaking in $N=2$
  Supersymmetric QCD,''} \hepth{9408099}}

\lref\SWSIXD{N. Seiberg and E. Witten,
  {\it ``Comments On String Dynamics In Six-Dimensions,''}
  \hepth{9603003}}

\lref\VafaFT{C. Vafa,
  {\it ``Evidence For F-Theory,''} \hepth{9602022}}

\lref\BDS{T. Banks, M.R. Douglas and N. Seiberg,
  {\it ``Probing F-theory With Branes,''}
  \hepth{9605199}}

\lref\SeiPRB{N. Seiberg,
  {\it ``IR Dynamics on Branes and Space-Time Geometry,''}
  \hepth{9606017}}

\lref\ABKSS{O. Aharony, M. Berkooz, S. Kachru, N. Seiberg and
  E. Silverstein,
  {\it ``M(atrix) description of $(2,0)$ theories,''} \hepth{9707079}}

\lref\Lowe{D. Lowe,
  {\it ``$E_8 \times E_8$ Small Instantons in Matrix Theory,''}
  \hepth{9709015}}

\lref\BFSS{T. Banks, W. Fischler, S.H. Shenker and L. Susskind,
  {\it ``M Theory As A Matrix Model: A Conjecture,''} \hepth{9610043}}

\lref\KSAB{O. Aharony, M. Berkooz, S. Kachru and E. Silverstein,
  {\it ``M(atrix) description of $(1,0)$ theories,''} \hepth{9709118}}

\lref\WitQHB{E. Witten,
  {\it ``On The Conformal Field Theory Of The Higgs Branch,''}
  \hepth{9707093}}

\lref\SusANO{L. Susskind,
  {\it ``Another Conjecture about M(atrix) Theory,''} \hepth{9704080}}

\lref\BluInt{J.D. Blum and K. Intriligator,
  {\it ``New Phases of String Theory and 6d RG Fixed Points via Branes at
  Orbifold Singularities,''} \hepth{9705044}}

\lref\IntNEW{K. Intriligator,
  {\it ``New String Theories in Six Dimensions via Branes at Orbifold
  Singularities,''} \hepth{9708117}}

\lref\WitPMF{E. Witten,
  {``Phase Transitions In M-Theory And F-Theory,''}
  \hepth{9603150}}

\lref\SeiFIV{N. Seiberg,
  {\it ``Five Dimensional SUSY Field Theories, Non-trivial Fixed Points
  AND String Dynamics,''} \hepth{9608111}}

\lref\Conrad{J.O. Conrad,
  {\it ``Brane Tensions and Coupling Constants from within M-Theory,''}
  \hepth{9708031}}

\lref\Leigh{R.G. Leigh, \mpl{4}{89}{2767}.}

\lref\Polch{J. Polchinski,
  {\it ``TASI Lectures on D-Branes,''} \hepth{9611050}}

\lref\GubKle{S.S. Gubser and I.R. Klebanov,
  {\it ``Absorption by Branes and Schwinger Terms
  in the World Volume Theory,''} \hepth{9708005}}

\lref\KV{S. Kachru and C. Vafa,
  {\it ``Exact Results For $N=2$ Compactifications
  Of Heterotic Strings,''} \hepth{9505105}}

\lref\LeeTAP{S. Lee, to appear.}

%%% ------------------------- CUT HERE ---------------------------------%

% ===================================================================== %
% Introduction
% ===================================================================== %
\newsec{Introduction}

In the past two years, many examples of nontrivial IR fixed
points in various dimensions have emerged.
Some of the most exciting ones are the 5+1D chiral theories.
The first of such theories with $\SUSY{(2,0)}$ SUSY has been discovered
in \WitCOM\ as a sector of type-IIB compactified on an $A_1$
singularity. A dual realization was found in \StrOPN\
as the low-energy description of two 5-branes of M-theory.
Another theory of this kind arises as an M-theory
5-brane approaches the 9-brane \refs{\GanHan,\SWSIXD}.
When the distance between
the 5-brane and 9-brane is zero, the low-energy is described
by a nontrivial 5+1D fixed point. This theory is chiral
with $\SUSY{(1,0)}$ SUSY and a global $E_8$ symmetry.
In \refs{\SWSIXD,\WitV} more examples of $\SUSY{(1,0)}$
theories have been given.
We will use the terminology of \WitV\ and call the $E_8$ theory
$V_1$.
Many other 5+1D theories have been recently
constructed in \refs{\BluInt,\IntNEW}.

M(atrix)-theory \refs{\BFSS,\SusANO} has sparked
a lot of progress for the 6D cases \refs{\ABKSS,\WitQHB,\Lowe,\KSAB}.
Nevertheless, these theories, which have no
coupling-constant around which to expand, are still mysterious.
The pieces of information that are known concern the BPS
spectrum and the low-energy effective actions in 6D,
and in 5D and 4D after toroidal compactification.
It is also known that $\SUSY{4}$ four-dimensional SYM as well as
$\SUSY{2}$ four-dimensional QCD with $N_f=0\dots 4$ flavors,
can be obtained
by appropriate limits of compactification of the 6D theories
on a torus \refs{\WitCOM,\GanTOR,\GMS}.

String theory is a powerful tool to study such theories.
The idea is to identify a dual description such that quantum
corrections of the original theory appear at the classical level
of the dual \KV.
The toroidal compactification of the
$\SUSY{(1,0)}$ 6D theory (and hence 4D $\SUSY{2}$ QCD) can be studied
using the brane-probe technique discovered in \refs{\SeiPRB,\BDS}.
The world-volume theory on a brane probe in a heterotic
string vacuum (which is quantum mechanically corrected)
is mapped by duality to a world-volume theory on a brane inside a curved
background which is not quantum mechanically corrected.
This allows one to determine the low-energy behavior in 4D.
At the origin of the moduli space
one obtains an IR fixed point with $E_8$ global symmetry.

The purpose of the present work is to extract information about
the local operators of such theories.
The $E_8$ theory $V_1$ 
has a local $E_8$ current $j^a_\u(x)$ ($a=1\dots 248$
and $\u=0\dots 5$).
We will be interested in the correlator  $\ev{j^a_\u(x) j^b_\v(0)}$.
The strategy will be to couple the theory to a weakly coupled $E_8$
gauge theory and calculate the effect of $V_1$ on the $E_8$ 
coupling constant.
We will study the question both for the 5+1D theory and for
the 3+1D conformal theories.
We will present three methods for evaluating the correlator.
The first method is purely field-theoretic and applies to the
3+1D theories. Deforming the theory with a relevant operator
one can flow to the IR where a field-theoretic
description of $SU(2)$ or $U(1)$ with several quarks \GMS\ can be
found.
This will allow us to determine the correlator as a function
on the moduli space.
From this function we can deduce the high-energy behavior
of the correlator and find out how many copies of the $E_8$
theory can be gauged with an $E_8$ SYM before breaking 
asymptotic freedom.

The other two methods for determining the correlators involve 
M-theory and F-theory.
The gravitational field of a 5-brane of M-theory which
is close to a 9-brane changes
the local metric on the 9-brane. After compactification
on a large $K3$ this implies that the volume of the $K3$ at the 
position of the 9-brane is affected by the distance from the 5-brane
(see \WitNWT). This can be interpreted as a dependence of the $E_8$
coupling constant on the VEV which specifies the position of the 5-brane.
From this fact we can extract the current correlator.
The third method involves the F-theory \VafaFT\ realization of the $E_8$
theory \refs{\SWSIXD,\WitPMF,\MVII}.
The $V_1$ theory is obtained in F-theory compactifications on a 
3-fold by blowing up a point in the (two complex dimensional) base.
By studying the effect of the size of the blow-up on the size of
the 7-brane locus we can again determine the dependence of the 
$E_8$ coupling constant on the VEV.

The paper is organized as follows.
In section (2) we give a brief review of the 6D and 4D theories.
In section (3) we calculate the current-current correlators in 3+1D
using field theory arguments and we argue that 10 copies of the
$E_8$ theory can be coupled to a gauge field.
In section (4) we study the effect of a 5-brane on the volume of
a 9-brane in M-theory and deduce the correlator from this setting.
In section (5) we present the F-theory derivation.
In section (6) we conclude with remarks and observations.
We have made an effort to make these notes more or less self-contained.

%%% ------------------------- CUT HERE ---------------------------------%

% ===================================================================== %
% Section (2): Review of the 6D and 4D theories
% ===================================================================== %
\newsec{Review of the 6D and 4D theories}

In this paper we are going to study conformal theories
with 8 supersymmetries in 5+1D and in 3+1D.
The theories that we are going to consider in 3+1D have a moduli space
parameterized by a single complex scalar. At the generic point in
the moduli space the super-conformal symmetry is spontaneously
broken and the low-energy description is a single $U(1)$
vector-multiplet whose interaction is given by a certain Seiberg-Witten
curve.  We will denote the coordinate on the moduli space by $u$.
We choose it such that $u=0$ is the point where the super-conformal
symmetry is unbroken.
Most of the theories that we will discuss also have a Higgs branch
emanating from the point $u=0$, but we will not discuss that branch
in this paper.

The Seiberg-Witten curves for the theories will be of the form
$$
y^2 = x^3 - f(u) x - g(u),
$$
where $f(u)$ and $g(u)$ are certain specific polynomials.
An example of such a super-conformal theory is given by
$\SUSY{2}$ $SU(2)$ QCD with $N_f = 4$ massless quarks.
The SW curve for this theory is given by \SWQCD:
$$
y^2 = x^3 - a u^2 x - b u^3.
$$
This theory also has a global $SO(8)$ symmetry under which the
vector-multiplet is a singlet \SWQCD.

It turns out that there are more exotic super-conformal theories
in 3+1D with the exceptional groups $E_6$,$E_7$,$E_8$
as global symmetries \refs{\MNI,\GanTOR,\MNII,\GMS,\MNWI,\MNWII}.
These theories do not have a known construction in terms of
the IR fixed point of some known field-theory.
However, they can be constructed from the low-energy degrees
of freedom of type-IIA compactifications on a singular Calabi-Yau \GMS.
The list of known super-conformal theories can be characterized
partly by the type of global symmetry of the theory \GMS.

Another (related) method of constructing such theories is to
start with the exotic 5+1D theory of small $E_8$ instantons
\refs{\GanHan,\SWSIXD}.
This theory has $\SUSY{(1,0)}$ supersymmetry and a ``Coulomb-branch''
where the low-energy description is a single tensor multiplet
comprising an anti-self-dual 2-form $B_{\u\v}^{(-)}$,
a scalar $\phi$ and fermions.
The VEV of the scalar $\phi$ parameterizes the
moduli space $\MR{1}/\BZ_2$.
At a generic point in moduli space the super-conformal symmetry is
spontaneously broken.
The origin of moduli space $\phi=0$ is the point where the
super-conformal symmetry is restored.
This $E_8$ theory also has a global $E_8$ symmetry and a Higgs
branch on which the $E_8$ symmetry is spontaneously broken.
This Higgs branch emanates from the $\phi=0$ point, but we will not
discuss it in this paper.
By compactifying the 5+1D theory on a $\MT{2}$ one obtains
a 3+1D Seiberg-Witten curve of the form \refs{\GanTOR,\GMS}:
$$
y^2 = x^3 - a u^4 x - b u^5 - c u^6.
$$
where $a,b,c$ are constants which depend on the compactification
parameters.
After compactification the theory is no longer super-conformal.
Its scale is set by the size of the $\MT{2}$.
However, from this construction one can extract a 3+1D super-conformal
theory. We define it as the IR limit of the theory at the origin
of moduli space $u=0$ \GMS. It has a Coulomb branch with a SW curve of
the form
$$
y^2 = x^3 - b u^5.
$$
One can also compactify with nontrivial $E_8$ boundary conditions
along the $\MT{2}$ \refs{\GanTOR,\GMS}.
In this way one gets a more general SW curve
$$
y^2 = x^3 - f_4(u) x - g_6(u).
$$
where $f_4$ and $g_6$ are polynomials of degrees
$4$ and $6$ respectively.
For certain values of the $E_8$ boundary conditions (``Wilson-lines'')
there are points in the moduli space of the  resulting 3+1D theory
where the IR limit is an interacting  IR super-conformal
theory.
 In this way one can get theories with $E_6$,$E_7$ and $E_8$
global symmetries as well as the fixed points which can be obtained
in QCD \GMS.

In what follows we will need the relation between $u$ and
the parameters of the $V_1$ theory in the limit $u\rightarrow\infty$.
It can be argued \GMS\ that in this limit
$$
u \sim e^{T A + i B}
$$
where $A$ is the area of the $\MT{2}$, $T$ is the tension of the
BPS strings of the 5+1D theory (proportional to the VEV of the
scalar $\phi$ of the tensor multiplet) and $B$ is the integral of the
anti-self-dual 2-form $B_{\u\v}^{(-)}$ over $\MT{2}$.

%------------------------------------------------------------------%
% Casimirs etc.
%------------------------------------------------------------------%
\subsec{Currents in gauge theories}

In this section we will set our conventions and review some elementary
facts from gauge theory. Let us consider a simple gauge group G.
Let $T^a$ ($a=1\dots \dim G$) be a set of generators with 
$$
\Trr{\rm adj}{T^a T^b}= C_2(G)\, \delta^{ab},
$$
where  $C_2(G)$ is the quadratic 
Casimir of the group. Define the nonabelian field strength
$F=F^a T^a$ and let the action be given by 
$$
S= \frac1{4g^2} \int d^4x F^a F^a - \int j^a(x) A^a(x) d^4x + \cdots
$$
where $j^a$ is a current coupled to the gauge field.
The current would come from some matter coupled to the gauge theory.
We are interested in the current-current correlation function,
$$
 \ev{j^a_{\mu}(q) j^b_{\nu}(p)} 
= (2\pi)^4 \delta(p+q)\,\ev{j^a_{\mu}(q) j^b_{\nu}(-q)}
$$
The Fouri\'er transform depends
on a cut-off, $\Lambda$ and in 3+1D 
the part including the cut-off will
be of the form
$$
\ev{j^a_{\mu}(q) j^b_{\nu}(-q)}=
c\, \delta^{ab}(q^2\eta_{\u\v} - q_\u q_\v)\times
\cases{
\logx{\frac{\Lambda}{m}} & if $|q|\ll m$ \cr
\logx{\frac{\Lambda}{|q|}} & if $|q| \gg m$ \cr
}
$$
where $m$ is a typical mass scale of the matter theory and
$c$ is some constant.
The coupling constant at a scale $\mu$ will run for $\mu \gg m$ as
$$
\frac1{g^2(\mu)} = \frac1{g^2(\Lambda)} - c \, \logx{\frac{\Lambda}{\mu}}
$$
or in other words the $\beta$-function is
$$
\beta = \mu \frac{dg}{d\mu} = - \frac{c}{2} g^3.
$$
For $\mu \ll m$,the coupling will be fixed at
$$
  \frac1{g^2({\mu})}= \frac1{g^2(\Lambda)} - c\, \logx{\frac{\Lambda}{m}}.
$$
The value of $c$ for bosons and fermions can be calculated by standard
field theoretic methods. Let $\rep{R}$ be a representation of $G$ and define
$$
\trr{\rep{R}}{T^a T^b} = C(\rep{R})\, \delta^{ab}.
$$
For complex bosons in representation $\rep{R}$
$$
c = - \frac{C(\rep{R})}{24 \pi^2},
$$
and for Dirac fermions in $\rep{R}$
$$
c = - \frac{4 C(\rep{R})}{24 \pi^2}.
$$
If we think of the gluons as a source of current they have a value of $c$,
which is
$$
c =  \frac{11 C_2(G)}{24 \pi^2}.
$$
Collecting all this we get the standard formula for the $\beta$-function
of a gauge theory 
$$
\beta = \frac{-11 C_2(G) + 4 C({\rm Dirac\ fermion\ repr.}) +
        C({\rm complex\ boson\ repr.})} {48 \pi^2} g^3.
$$
For $\SUSY{2}$ theories we have vector-multiplets in the adjoint
representation.  A vector-multiplet has a value of $c$ which is
\eqn\ctwog{
c = \frac{C_2(G)}{4 \pi^2}.
}
Hyper-multiplets in representation $\rep{R}$ have a value of $c$ which is
$$
c = - \frac{C(\rep{R})}{4 \pi^2}.
$$
We see especially that to saturate the $\beta$-function we need enough 
hyper-multiplets so $C(\rep{R})=C_2(G)$.
For a simple group G and a representation $\rep{R}$
 the ratio $\frac{C_2(G)}{C(\rep{R})}$
is a calculable number. In this paper we will only need the result for SO(N)
and the fundamental representation where
$$
\frac{C_2(SO(N))}{C({\rm fundamental\ of\ } SO(N))} = N-2.
$$
We see that $N-2$ fundamental hyper-multiplets of $SO(N)$
saturate the $\beta$-function.

%%% ------------------------- CUT HERE ---------------------------------%

% ===================================================================== %
% Section (3): Field theory derivation
% ===================================================================== %
\newsec{The current-current correlator -- field theory derivation}

In this section we will derive the form of the $E_8$ current-current
correlator for the $E_8$ conformal theory and as a result
we will argue that in 4D one can couple up to 10 copies of the $E_8$
theory to a $\SUSY{2}$  $E_8$ Yang-Mills gauge theory.

We start with the $E_8$ conformal theory in 4 dimensions whose Seiberg-Witten
curve is given
by \GMS:
\eqn\swcee{
y^2 = x^3 + u^5,
}
$u$ parameterizing the moduli space of the Coulomb branch.
We are looking for an expression of the form 
\eqn\fqul{
\ev{j^a_\u (q) j^b_\v(-q)} = (q^2 \eta_{\u\v} - q_\u q_\v ) \delta^{ab}
f(q^2, u, \Lambda),\qquad a,b=1\dots 248,
}
where $q$ is the momentum and $\Lambda$ is some fixed UV-cutoff.
This UV-cutoff is not physical. It is just an artifact of the
 Fouri\'er transform. The space-time correlator 
$\ev{j^a_\u (x) j^b_\v(y)}$ does not require a cutoff.

Let us first calculate the dimension of $u$,
using the technique of \APSW. From \swcee\ we get the equations 
$$
{\rm dim}\lbr x \rbr = {5\over 3}{\rm dim}\lbr u\rbr,\qquad
{\rm dim}\lbr y \rbr = {5\over 2}{\rm dim}\lbr u\rbr.
$$
Since $a \sim \int {{dx}\over {y}}\wdg du$ has dimensions of mass
we find 
$$
u  \sim {\rm Mass}^6.
$$

To determine the form of $f$ in \fqul\ for $q^2 = 0$
we can couple the $E_8$ SCFT to a weakly coupled $E_8$ gauge field
and ask how the $E_8$ coupling constant changes as a function of $u$.
When the $E_8$ coupling constant is very small the coupling
 does not change the curve \swcee\ by much.
For a generic value of $u$ the massless modes of the $E_8$ SCFT
are neutral under the global $E_8$ and the charged matter 
has a typical energy of order $u^{1/6}$.
The $\ev{jj}$ correlator will modify the low energy $E_8$ coupling constant
to the form
$$
{1\over {g(u)^2}} = {1\over {(g_0)^2}} + f(q^2 = 0, u, \Lambda),
$$
where $g_0$ is the bare coupling constant.
On the other hand, standard renormalization arguments require
that it should be possible to re-absorb the $\Lambda$ dependence in
the bare coupling constant. Thus, dimensional analysis restricts
the form of $f(0,u,\Lambda)$ to 
$$
f(0,u,\Lambda) = c\, \logx{ {\Lambda \over {|u|^{1/6}}} },
$$
where $c$ is a constant as discussed in section (2.1).

Before we determine $c$, let us see how many copies of the 
$E_8$ SCFT can be consistently coupled to an $E_8$ gauge field
without ruining asymptotic freedom.
Unlike the previous discussion, this is a question about the UV
behaviour of the theory. Thus we can fix $u$ and take 
$|q|\gg u^{1/6}$. This means that we are interested in
the value of $f$ when $u=0$ i.e. $f(q^2,0,\Lambda)$.
Since the $\Lambda$ dependence of $f$ must still be the same 
as before, using dimensional analysis we conclude that 
\eqn\contha{
f(q^2 ,0,\Lambda) = \half c\, \logx{ {{\Lambda^2} \over {q^2}} }.
}

To determine $c$ we deform the theory by adding a relevant
operator to its (unknown) Lagrangian such as to break the
global $E_8$ symmetry down to $D_4$ ($SO(8)$) by putting Wilson lines
on the torus (see also \refs{\MNI,\MNII,\GMS,\MNWI,\MNWII}).
The advantage is that the $D_4$ conformal fixed point 
can be analyzed  in standard field-theory. It is the IR free
theory of $SU(2)$ coupled to 4 massless quarks \GMS.

The deformation to a $D_4$ curve is given by
\eqn\swcdf{
y^2 = x^3 + u^5 + \lam^8 u^2 x + \a \lam^{12} u^3,
}
where $\lam$ is a parameter with dimensions of mass,
and the form of the deformation was extracted from the 
elliptic-singularity-type tables of \MVII.

The discriminant of \swcee\ had a single zero of order
10 at the origin. This $E_8$-type singularity has split into 
five singularities in \swcdf. One is a $D_4$-type singularity
at $u=0$ and the other four are $A_0$-type singularities
(i.e. can be modeled by a $U(1)$ with one massless electron).

The global $E_8$ of the original theory has been broken by
the operators to a global $SO(8)$.
For the theory with curve \swcdf\ we can ask what is
\eqn\jjsof{
\ev{j_\u^A(q) j_\v^B(-q)},\qquad A,B = 1\dots 28,
}
where $A,B$ are $SO(8)$ indices.

The curve near $u = 0$ looks like
$$
y^2 = x^3 + \lam^8 u^2 x + \a \lam^{12} u^3.
$$
Now we rescale
$$
u = \lam^4 \wbu,\qquad
x = \lam^8 \wbx,\qquad
y = \lam^{12} \wby.
$$
This preserves the form ${{dx}\over {y}}\wdg du$ and the curve looks
like
$$
\wby^2 = \wbx^3 + \wbu^2 \wbx + \a \wbu^3.
$$
This is the curve which describes the low-energy near the $D_4$ singularity.
On the other hand, the low-energy of the $D_4$ singularity can be described
by $SU(2)$ with  4 quarks.
The moduli parameter $\wbu$ should be identified with the VEV of the
$U(1)$ such that the mass of the quarks is proportional to $m=\wbu^{1/2}$
(by dimensional analysis).

Near the $D_4$ singularity we can use field theory to calculate
$\ev{j^A j^B}$ where $A,B$ are $SO(8)$ indices.
The $D_4$ theory contains one hyper-multiplet in the fundamental of $SO(8)$.
 From section (2.1) we see that such a theory has 
\eqn\abc{\eqalign{
 \ev{j^A_{\mu}(q)  j^B_{\nu}(-q)} &=  - \frac{C({\rm fund.})}{4 \pi^2}   
\delta^{AB}    
 (q^2 \eta_{\mu\nu} -q_{\mu}q_{\nu})\,
      \log \left| \frac{\Lambda}{\wbu^{1/2}} \right| \cr
  &= 
 - \frac{3 C({\rm fund.})}{4 \pi^2}   
\delta^{AB}    
 (q^2 \eta_{\mu\nu} -q_{\mu}q_{\nu})\,
  \log \left| \frac{(\Lambda \lam^2)^{1/3}}{u^{1/6}} \right| \cr
}}
This equation is valid for small $u$. By holomorphy it must have the same
 functional form for large $u$.
This is because \abc\ for $q=0$, is part of a holomorphic expression.
If we couple the theory to a weakly coupled $SO(8)$ gauge
field then \abc\ will be the correction to the $SO(8)$ coupling
constant and the imaginary part of the $\log$ will be the correction
to the $\theta$-angle. Thus the coupling constant together with
the $\theta$-angle are holomorphic in $u$.
The exponential of the $\log$ in \abc\ has to be a single valued
function of $u$. It can have a singularity (a zero or a pole)
only when $SO(8)$-charged
matter becomes massless. This never happens when $u\neq 0$.
The other four $A_0$ singularities in the moduli space
correspond to singlets of the $SO(8)$ which become massless.
Furthermore, the physical behaviour at infinity restricts 
\abc\ to diverge at most logarithmically in $u$.
It follows that the form \abc\ has to be valid for all $u$.
For $u\rightarrow\infty$ the perturbation in \swcdf\ is negligible
because $\lam$ is small compared to the scale set by $u$.
Thus, we can read off \jjsof\ from \fqul.
%%% Here we argued above that $\ev{j^A j^B}$
%%% was the same in the $D_4$ theory and in the $E_8$ theory.
We conclude that for the $E_8$ theory
\eqn\ghi{\eqalign{
\ev{j^a_{\mu}(q)  j^b_{\nu}(-q)} &=
   - \frac{3 C({\rm fund.})}{4 \pi^2}   
\delta^{ab}
 (q^2 \eta_{\mu\nu} -q_{\mu}q_{\nu})\,
  \log\left|\frac{(\Lambda \lam^2)^{1/3}}
  {u^{1/6}}\right|. \cr
}}
This means that the value of $c$ in \contha\ is 
\eqn\jkl{\eqalign{
c &= - 3 \frac{C({\rm fund.\ of\ SO(8)})}{4 \pi^2} \cr
  &= - 3 \frac{C({\rm fund.\ of\ SO(8)})}{C_2(SO(8))}
       \frac{C_2(SO(8))}{C_2(E_8)}
        \frac{C_2(E_8)}{4 \pi^2} \cr
  &= -  3 \times\inv{6} \times\frac{6}{30}  \frac{C_2(E_8)}{4 \pi^2} \cr
  &= -  \inv{10} \times\frac{C_2(E_8)}{4 \pi^2}.\cr
}}
Comparing with \ctwog\ we see that 10 copies of the $E_8$ SCFT
can be coupled to an $E_8$ SYM.

%------------------------------------------------------------------%
% D_5
%------------------------------------------------------------------%
We can similarly deform the theory by adding a relevant
operator to its (unknown) Lagrangian such as to break the
global $E_8$ symmetry down to $D_5$ ($SO(10)$) (see also
\refs{\MNI,\MNII,\GMS,\MNWI,\MNWII}).
The vicinity of the $D_5$ point can be represented
by the IR free theory of $SU(2)$ coupled to 5 massless quarks \GMS.

The deformation to a $D_5$ curve is given by
\eqn\swcdf{
y^2 = x^3 + u^5 + (2 \lam^{12} u^3 + \a \lam^6 u^4 - 3\lam^8 u^2 x),
}
where $\lam$ is a parameter with dimensions of mass, $\a$
is a dimensionless parameter
and the form of the deformation was extracted from the 
elliptic-singularity-type tables of \MVII.

The discriminant of \swcee\ had a single zero of order
10 at the origin. This $E_8$-type singularity has split into 
four singularities in \swcdf. One is a $D_5$-type singularity
at $u=0$ and the other three are $A_0$-type singularities
(i.e. can be modeled by a $U(1)$ with one massless electron).

The global $E_8$ of the original theory has been broken by
the operators to a global $SO(10)$.
For the theory with curve \swcdf\ we can ask what is
\eqn\jjsof{
\ev{j_\u^A(q) j_\v^B(-q)},\qquad A,B = 1\dots 45,
}
where $A,B$ are $SO(10)$ indices.

%%% For $u\rightarrow\infty$ the perturbation in \swcdf\ is negligible
%%% because $\lam$ is small compared to the scale set by $u$.
%%% Thus, we can read off the \jjsof\ from \fqul.
%%% On the other hand,
We can calculate \jjsof\ by modeling the
vicinity of $u=0$ as $SU(2)$ with 5 quarks.
For this purpose we need to determine the relation
between our $u$ and the field-theoretic  $\wu = \trp{\phi^2}$
(where $\phi$ is the $SU(2)$ field).
%%% $$
%%% x\rightarrow x + \lam^4 u
%%% $$
%%% The curve near $u = 0$ looks like
%%% $$
%%% %%% y^2 = x^3 - 3 \wbu^2 x + 2 \wbu^3 + \a \lam^{-10}\wbu^4
%%% %%%       + 3 \wbu x^2 + 3 \wbu^2 x + \wbu^3 - 3 \wbu^3
%%% y^2 = x^3 + 3 \lam^4 u x^2 + \a \lam^6 u^4
%%% $$

This can be done by calculating the coupling constant $\tau$
near $u=0$.
From the curve \swcdf\ we find
$$
\tau \sim -{1\over {2\pi i}}\,
\log \left({{\a u}\over {\lam^6}}\right) + \Ol{1}.
$$
On the other hand, the 1-loop field-theory result is:
$$
\tau \sim -{1\over {2\pi i}}\,
\log \left({{\trp{\phi^2}}\over {\Lambda^2}}\right) + \Ol{1}.
$$
where $\tau$ here is the coupling constant of the unbroken
$U(1)\subset SU(2)$ in the conventions of \SWQCD.
Thus we may identify
$$
u = \mu_0\, \trp{\phi^2}
$$
where $\u_0$ is some dimensionful constant.
Now we can couple the field theory to a weakly coupled $SO(10)$
gauge field and continue as before. We get the same result.

%%% ------------------------- CUT HERE ---------------------------------%

% ===================================================================== %
% Section (4) M-theory calculation
% ===================================================================== %
\newsec{Derivation from M-theory}

The system of the $(1,0)$ $E_8$ theory ($V_1$) coupled to $E_8$ SYM
can be realized in M-theory as a 5-brane which is close to the 9-brane.
The modes of the $V_1$ theory come from the 5-brane bulk and from
membranes stretched between the 5-brane and 9-brane while the $E_8$
SYM comes from the 9-brane  bulk. Let us compactify on $K3\times\MT{2}$.

The effect that we are trying to study corresponds to the following
question.
The gravitational field of the 5-brane affects the metric at the
position of the 9-brane. Thus, as we change the distance of the 5-brane
from the 9-brane the volume of the $K3$ changes as a function of $x$
\WitNWT.
The volume of $K3\times\MT{2}$ is related to 
the 3+1D $E_8$ coupling constant.
In field-theory, 
this is interpreted as a running of the 
$E_8$ coupling constant as a result of the change of the VEV of the 
$V_1$ theory.

%%% We now set out to calculate this effect.
We apply the general setting and formulae of \WitNWT\
to the case where
the distance of the 5-brane from the 9-brane is much smaller
than the compactification scale of $K3\times\MT{2}$
and calculate the effect.

We must also mention that the after compactification of the
system of a 5-brane and 9-brane on $\MS{1}$ we get 4-branes near
8-branes. This setting has been studied in \SeiFIV, in the context
of brane probes, where a related effect is observed. 
The position of the probe affects the value of a classical field,
in that case the dilaton, which is then re-interpreted as a 1-loop
effect in field theory.
In fact, the relation between the classical
supergravity calculation and the 1-loop field-theory
calculation follows from perturbative string-theory.
The 1-loop result is a loop of DD strings connecting the 4-brane
to the 8-brane while the classical supergravity result is the same diagram
viewed from the closed string channel.

%------------------------------------------------------------------%
% Geometrical setup
%------------------------------------------------------------------%
\subsec{Geometrical setup and review}

In this section we will examine the theory of a 5-brane in 
M-theory on $\MR{5,1} \times K3 \times \MS{1}/\BZ_2$ and review
some relevant facts from \HWII\ and \WitNWT.

The geometric setup is as follows.
The coordinates $ (x^1 , x^2, ... ,x^6) $ parameterize $\MR{5,1}$,
$(x^7 , x^8 , x^9  ,x^{10} )$
parameterize K3 and finally $x^{11}$ parameterize $\MS{1}/\BZ_2$.
All 5-branes have their world-volume along $\MR{5,1}$ and  are located
at a point
in $K3 \times \MS{1}/\BZ_2$. All configurations will be defined on the
whole ${\bf S^1}$ and are symmetric under the $\BZ_2$ (working
``upstairs'' -- in the terminology of \HWII). This means,
for example, that every time there is a 5-brane between the two fixed planes
of the $\BZ_2$ there is also a mirror 5-brane. There would be an
equivalent formulation (``downstairs'') where configurations
were only defined on the interval between the two ``ends of the world''.

We know that M-theory on $\MR{9,1} \times \MS{1}/\BZ_2$ is heterotic
$E_8 \times E_8$
with one $E_8$ theory living on each fixed plane of the $\BZ_2$.
If we compactify
this theory on K3 we need to supply a total of 24 instantons and 5-branes.
The theory we are interested in is a single
5-brane coupled to an $E_8$ gauge theory.
To achieve this we need to have no instantons in one of the $E_8$
theories and one 5-brane close to this ``end of the world''.
The remaining 23 instantons and 5-branes must therefore be either
instantons in the other ``end of the world'' or 5-branes in the bulk.
Our configuration is shown in the figure.
\iffigs
\midinsert
$$\vbox{\centerline{\epsfxsize=4in\epsfbox{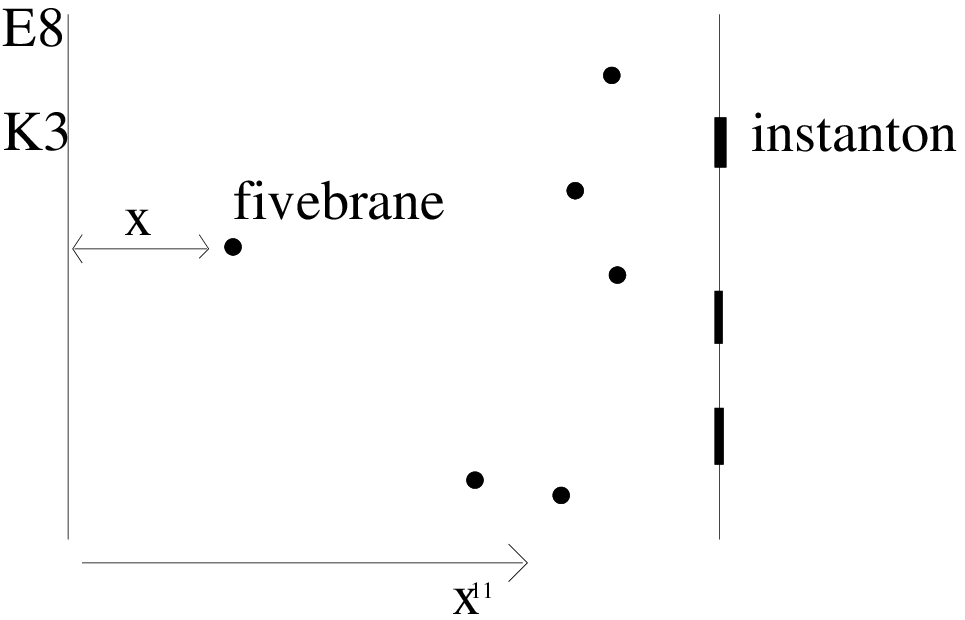}}
{Figure 1: The geometry of our setup.
We are considering M-theory on $\MR{5,1} \times K3 \times \MS{1}/\BZ_2$.
The horizontal direction is along the circle. Only half of the circle
is shown. There is a mirror image not shown here. The vertical direction
is along K3 and $\MR{5,1}$ is not shown.
There are no instantons at $x^{11}=0$
so the gauge group is $E_8$. The dots represent 5-branes.
The relevant 5-brane is at a distance $x$, from ``the end of the world''
at $x^{11}=0$.}
}$$
\endinsert
\fi

In the 6-dimensional description the distance of the 5-brane from the
``end of the world,'' $x$, is a modulus.
The effective gauge coupling of the $E_8$ depends on $x$.
 From the 6-dimensional point of view certain degrees of freedom connected
to the 5-brane act as matter coupled to the $E_8$ gauge field.
Since the couplings and masses of this matter depend on $x$,
the low energy effective $E_8$ gauge coupling, $g$, will depend on $x$.
Here we will calculate the $x$-dependence of $g$ from M-theory or more
precisely from 11-dimensional supergravity. For supergravity to be
applicable all distances involved in the problem need to be much
bigger than the 11-dimensional Planck scale. This means especially
that ${\rm Vol}(K3) \gg l_{Planck}^4$.
Furthermore we are interested in the behaviour of the theory when it is
close to the point with tensionless strings or equivalently with a zero
size instanton, which is $x = 0$.
To be in that situation we take $x \ll vol(K3)^{\inv{4}}$.
 The $x$-dependence of the 6-dimensional gauge coupling $g$,
comes about because the volume of the K3 at $X^{11}=0$ depends on $x$.

To calculate $g$ we need to find the form of the metric as a function
of $x$. Luckily most of this has been done in \WitNWT.
 We will now review the relevant facts from that paper.
The gauge and gravitational part of the action of M-theory on
$\MS{1} / \BZ_2$ takes the form
\eqn\maction{ 
S = - \inv{4\kappa^2} \int_{M^{11}} d^{11}x \sqrt{g}R\, 
- \sum_{i=1,2} \inv{16\pi(4\pi \kappa^2)^{2/3}} 
\int_{M_i^{10}}  d^{10}x \sqrt{g} \;  tr F_i^2           
}
where $\kappa$ is the 11-dimensional gravitational coupling and
 $tr$ is $\inv{30}$ of the trace in the adjoint representation of $E_8$.
The integral is over $M^{11}=\MR{10} \times \MS{1}$
with all fields invariant under the $\BZ_2$.
There is a gauge kinetic term for each of the two
``ends of the world''  $M_i$  ($i=1,2$).
The coefficient in front of the gauge kinetic term can 
be found from purely 11-dimensional considerations \HWII. It can
also be found by comparison with 10-dimensional heterotic
string theory.
The coefficients in the action
are corrected as explained in \Conrad.\foot{We are very grateful
to Sangmin Lee for pointing our attention to this paper.
See also \LeeTAP.}
The unconventional normalization of the gravitational term
stems from the fact that we are working upstairs.
Downstairs the gravitational term would be multiplied by 2
and give the standard ${1\over {2\kappa^2}}$ (see \Conrad).

We now consider M-theory on $\MR{5,1} \times K3 \times \MS{1} / \BZ_2$.
The metric is determined by solving the equations for unbroken
supersymmetry.
The only fermionic field is the gravitino, so we only have to solve
$$ \delta \Psi_I = 0.$$
The transformation for $\Psi_I$ is
$$
 \delta \Psi_I = D_I \eta + \frac{\sqrt2}{288}
( \Gamma_{IJKLM} - 8 g_{IJ} \Gamma_{KLM}) G^{JKLM} \eta                   
$$ 
where $G$ is the 4-form field strength of M-theory.
$\eta$ is the supersymmetry transformation parameter.
Furthermore $G$ must obey the equation of motion
\eqn\eqofmot{
D^I G_{IJKL} =0
}
and the Bianchi identity
\eqn\kilder{
dG = \lbr {\rm sources}\rbr.
}

Here the sources are the instantons and curvature in the
``ends of the world'' and the 5-branes.
An important point is that the right-hand side of equation \kilder\
is proportional to $\kappa^{2/3}$. This means that to zeroth order
in $\kappa$ the equation for unbroken supersymmetry is 
$$
D^I \eta = 0.
$$
This is solved by the metric
$$
ds_{(0)}^2 = \eta_{\mu \nu} dx^{\mu} dx^{\nu} + g_{AB} dx^A dx^B
$$
$$
\mu,\nu=1\dots 6  \qquad     A,B=7\dots 11
$$
with $g_{AB}$ the product of a hyper-K\"ahler metric on $K3$ and the
standard metric with $g_{11,11}=1$ on $\MS{1}$.

However what we want is the exact solution including the $G$ field.
This is also where the 5-brane position, $x$, enters.
As explained in \WitNWT\ this solution can actually be found.

It is found as follows.
First one solves equation \eqofmot\ and equation \kilder\
in the zeroth order metric $ds_{(0)}^2$.
The only non-zero components of $G_{IJKL}$ are along the internal
5-dimensional manifold $K3 \times \MS{1}$. The solution can be written in
terms of a function $w$:
$$
G_{ABCD}=- \epsilon^{0}_{ABCDE} \partial^E w
$$
with $\epsilon^{0}$ the completely antisymmetric tensor in the metric
$ds_{(0)}^2$. $w$ solves 
\eqn\wtwo{
\nabla_0 w = {\rm sources}.
}
$w$ is to be thought of as the dual of the 3-form potential on the
5-dimensional manifold in the metric $ds_{(0)}^2$ and can be determined
up to an additive constant. The exact metric now turns out to be 
\eqn\exmet{
ds^2 = (c+2\sqrt2 w)^{-1/3} \eta_{\mu \nu}dx^{\mu} dx^{\nu} +
      (c+2\sqrt2 w)^{2/3} g_{AB} dx^A dx^B
}
where $c$ is a constant.

%------------------------------------------------------------------%
% The case at hand
%------------------------------------------------------------------%
\subsec{A 5-brane very close to the 9-brane}

After this review of how to obtain the metric let us go back to the
problem of finding the low energy gauge coupling. We need to dimensionally
reduce the gauge kinetic term for the relevant ``end of the world''.
 From equation \maction\ it is seen to be
$$
 - \inv{16\pi(4\pi \kappa^2)^{2/3}}  \int_{M_i^{10}}
 d^{10}x \sqrt{g} \;  tr F_i^2.
$$
First we need to consider the question of what is the metric in 6 dimensions.
Looking at equation \exmet\ we see that the metric in $\MR{5,1}$
has a factor that depends on the position in the internal manifold.
 From the 6-dimensional point of view this is unwanted and we should
take $\eta_{\mu\nu}$ as the metric. Of course we could replace
$\eta_{\mu\nu}$ by any other metric in 6 dimensions. The point is
just that the metric of the 6 directions of
$\MR{5,1}$ in the 11-dimensional metric \exmet\ is Weyl
rescaled by $(c+2\sqrt2 w)^{-1/3}$ compared to the metric used by
the 6-dimensional observer. In the dimensional reduction we should
take care to include the Weyl factor in both $\sqrt{g}$ and in the
contraction of indices in $F^2$. 
The dimensional reduction gives 
\eqn\dimred{\eqalign{
& - \inv{16\pi(4\pi \kappa^2)^{2/3}} \int_{K3} d^4x 
 \sqrt{(c+2\sqrt2 w)^{2/3}} (c+2\sqrt2 w)^{2/3} \sqrt{g_{AB}}
  \int  d^6x  \;  tr F^2  \cr
&=
 - \inv{16\pi(4\pi \kappa^2)^{2/3}} \int_{K3} d^4x 
 (c+2\sqrt2 w)  \sqrt{g_{AB}} \int  d^6x  \;  tr F^2.\cr
}}
We conclude that the gauge coupling, $g$, is given by
\eqn\gaugei{
\inv{4g^2} =
 \inv{16\pi(4\pi \kappa^2)^{2/3}} \int_{K3} d^4x 
 (c+2\sqrt2 w)  \sqrt{g_{AB}}
}
in conventions where $tr(T^a T^b) = \delta^{ab}$.
Since $tr$ is $1/30$ times the trace in the adjoint
we see that $C_2(E_8)=30$.  We can also
easily find the 6-dimensional gravitational constant from the
Einstein-Hilbert term in equation \maction:
\eqn\gravcoup{
\inv{2{\kappa_{(6)}}^2} =
 \inv{4\kappa^2}  \int_{K3 \times \MS{1}} d^5x 
 (c+2\sqrt2 w)  \sqrt{g_{AB}}.
}

Now we need to find the $x$-dependence of $w$. We need to solve
equation \wtwo\ for the configuration with a 5-brane in position $x$ and
a mirror 5-brane in position $-x$. The 5-branes, of course, also have
a definite position in $x^7,x^8,x^9,x^{10}$, which does not play a role.
Furthermore there are some other contributions to the total source
term which we do not need to worry
about to find the $x$-dependence of $w$. Let us calculate the difference,
$w_x - w_0$, between $w$ for the 5-brane at position $x$ and, say,
position $x^{11}=0$. Since equation \wtwo\ is linear in $w$ and
sources all the other sources drop out of $w_x-w_0$. Thus, $w_x-w_0$ is
given by solving equation \wtwo\ for a 5-brane at position
$x^{11}=x$, a 5-brane at position $x^{11}= -x$ and two anti 5-branes
at position $x^{11}=0$. The metric is $g_{AB}$ which is a product of
a hyper-K\"ahler metric for $K3$ and $(dx^{11})^2$ for $\MS{1}$.
The only obstacle to solving this is that the metric is complicated.
However we assume that the volume of $K3$ and the distance between
the two ``ends of the world'' is very large, so it is a good
approximation to solve the problem in flat $\MR{5}$ metric,
$g_{AB}=\delta_{AB}$.
This is so because $w$ falls off to zero away from the 5-branes.
The length scale of this fall-off is set by $\kappa$. The error
of setting $g_{AB}=\delta_{AB}$ is suppressed by the
ratio of the 11-dimensional Planck scale and the smallest length
scale in $g_{AB}$. This ratio is small in our setup.

Now from equation \gaugei\ we can calculate the $x$-dependence of $g$,
\eqn\xdepe{\eqalign{
   \inv{4g^2} = &\,
 \inv{16\pi(4\pi \kappa^2)^{2/3}} \int_{K3,x^{11}=0} d^4x 
 (c+2\sqrt2 w)  \sqrt{g_{AB}} \cr
= &\,
 \inv{16\pi(4\pi \kappa^2)^{2/3}} \int_{K3,x^{11}=0} d^4x 
 (c+2\sqrt2 w_0)  \sqrt{g_{AB}}  \cr
&+ \inv{16\pi(4\pi \kappa^2)^{2/3}} \int_{K3,x^{11}=0} d^4x 
 2\sqrt2 ( w_x - w_0 )  \sqrt{g_{AB}}. \cr
}}
The $x$-dependence is solely in the last term. As discussed above we
are only making a small mistake by setting $g_{AB}=\delta_{AB}$ in
this term. Here it is important, though, that the integrand goes to
zero so fast that almost the full contribution to the integral comes
from a small region in $K3$. We can take the derivative with respect
to $x$ to determine the $x$-dependence. 
\eqn\gdx{
 \frac{\partial}{\partial x} (  \inv{4g^2}) =
 \inv{16\pi(4\pi \kappa^2)^{2/3}} \int_{\MR{4}}  
 2\sqrt2 \partial_x w_x d^4y
}
To calculate this we just need to find $w_x$, which is the value
of $w$ in the background of two 5-branes at position $x^{11}=\pm x$.
Here we work in $\MR{5}$.
On a compact manifold one could not have branes alone,
since on a compact manifold the source terms have to add up to zero
cohomologically.

Let us first look at a single 5-brane in M-theory on $\MR{10,1}$.
We take the 5-brane to have position $x^7=x^8=x^9=x^{10}=x^{11}=0$.
As calculated in \WitNWT\ by the same method as used to find the
metric in our setup, the metric around the 5-brane is
\eqn\fivemet{
ds^2= (1 + \frac{2 \sqrt{2}q}{R^3})^{-1/3} \eta_{\mu\nu}dx^{\mu}dx^{\nu} +
      (1 + \frac{2 \sqrt{2}q}{R^3})^{2/3} \delta_{AB}dx^{A}dx^{B} 
}
where
$$
 \mu,\nu = 1\dots 6 \qquad  A,B=7\dots 11  
$$
where $R=\sqrt{x^A x_A}$ and $q$ is a constant. In other words,
$$
w = \frac{q}{R^3}
$$
for a 5-brane.
We will find the exact value of $q$ below.

It is now easy to find $w$ for a 5-brane at position $(0,0,0,0,x)$
and one at position $(0,0,0,0,-x)$,
\eqn\atpos{\eqalign{
w(y)&= \frac{q}{((y^7)^2+(y^8)^2 +(y^9)^2 
                      + (y^{10})^2 + (y^{11}-x)^2)^{3/2}} 
     \cr
    &+ \frac{q}{((y^7)^2+(y^8)^2 +(y^9)^2 
                      + (y^{10})^2 + (y^{11}+x)^2)^{3/2}} 
      \cr
}}
We then get 
$$
\int_{\MR{4},y^{11}=0} \partial_x w_x d^4y = 2 \cdot
\left(-\frac{3}{2}\right) \cdot
2 x q
\int_{\MR{4}} \inv{(y^2 + x^2)^{5/2}} d^4y
= -8 \pi^2 q
$$
 From equation \gdx\ we get  
$$
\frac{\partial}{\partial x} (  \inv{4g^2}) = 
  - \frac{\sqrt2 \pi q}{(4\pi \kappa^2)^{2/3}}
$$
and so we conclude
$$
   \inv{4g^2} =   \inv{4g_0^2}  
  - \frac{\sqrt2 \pi q}{(4\pi \kappa^2)^{2/3}} x
$$
We see that the $x$-dependence of $\inv{g^2}$ is linear.

Following the same line of logic we can readily find the $x$-dependence of
$\kappa_{(6)}$ from equation \gravcoup. Since we integrate over
$\MS{1}$ and the zeroth order metric is translation invariant in the
$\MS{1}$ direction, $\kappa_{(6)}$, will be exactly independent of $x$. 

We now see that it was very fortunate that the factor $(c+ 2\sqrt2 w)$
appeared to the first power in the  equations \gaugei\ and \gravcoup.
 Had that not been the case we would not have
gotten so simple results for the $x$-dependence of $g$ and $\kappa_6$.
Furthermore the constant $c$ would have entered the formulas which
would have been odd. We want to interpret the $x$-dependence of $g$ as
due to matter from the 5-brane theory coupled to the gauge theory.
The theory on the 5-brane has nothing to do with the number $c$
which should therefore not enter the formulas for the $x$-dependence.

To complete the calculation we need the value of $q$.
From equation
\fivemet\ we see that $q$ has dimension (-3). Since $q$ only
depends on $\kappa$, we conclude that $q$ is some number times
$\kappa^{2/3}$.
There are several ways of finding $q$.
One way is to calculate the tension of the 5-brane from the ADM
formula and then equate this to the known value for the 5-brane tension.

For the ADM formula 
in $D$ dimensions we use the gravitational action\foot{This
is the action in ``downstairs'' form, which is the relevant one
\Conrad.}
$$
S = - \inv{2\kappa^2} \int \sqrt{g}R\, d^D x
$$
and we consider a p-brane with metric
$$
ds^2=  H(r)    \eta_{\mu\nu}dx^{\mu}dx^{\nu} +
               K(r)         \delta_{AB}dx^{A}dx^{B} 
$$
where
$$
       \mu,\nu = 1,2,\dots,p+1 \qquad A,B=p+2,\dots,D
$$
and $r^2 = (x^{p+2})^2 + .... + (x^D)^2$, then the tension is given as
$$
T_p = - \inv{2\kappa^2} {\rm Vol}(S^{D-p-2})\; \lim_{r \rightarrow \infty}  
(p H'(r) + (D-p-2) K'(r))r^{D-p-2}
$$
Using this formula for our solution \fivemet\ the tension of
the 5-brane is
$$
T_5 = \frac{8 \sqrt2 \pi^2 q}{\kappa^2}
$$
The tension of the M 5-brane is
$$
T_5 = (\frac\pi2)^{1/3} \kappa^{-4/3}.
$$
Equating these two expressions for $T_5$ gives 
$$
q = \inv{8\; 2^{5/6} \pi^{5/3}} \kappa^{2/3}.
$$
We finally get
\eqn\gx{
 \inv{4g^2} =   \inv{4g_0^2}  - \inv{16 \cdot
2^{2/3} \pi^{4/3} \kappa^{2/3}} x.
}
Using the tension of a membrane in M-theory:
$$
T_2 = 2^{1/3} \pi^{2/3} \kappa^{-2/3},
$$
We can rewrite equation \gx\ as
\eqn\gxt{
\inv{g^2} =   \inv{g_0^2}  - \inv{8 \pi^{2}} x T_2.
}
The expression $x T_2$ is the tension of the strings in the six-dimensional
theory. This is because the membrane is stretched with one direction
along the 11th direction and two directions along $\MR{5,1}$.
Looking at the metric \fivemet\ we see that the Weyl factors exactly
drop out of the formula for the tension measured in the metric
$\eta_{\u\v}$ in $\MR{5,1}$.

Compactifying further down to 4 dimensions on a torus of area $A$
is straightforward
\eqn\gax{
\inv{g^2} =   \inv{g_0^2}  - \inv{8 \pi^{2}} A x T_2.
}
The gravitational coupling is independent of $x$ both in 6 and 4 dimensions.

As we reviewed in section (2),
this theory has a Seiberg-Witten curve given by
$$
y^2 = x^3 - \lambda_1^{-4} u^4 x - \alpha u^5 - \lambda_2^{-6} u^6
$$
where $\lambda_1$,$\lambda_2$ are parameters of mass dimension 1,
$\alpha$ is a parameter of dimension $(-6)$. The moduli space of the 
theory is parameterized by the dimensionless $u$.
Furthermore, for large $u$ the connection between $u$ and $x$ is 
$$
|u| = e^{AxT_2}.
$$
Substituting this value of $u$ in the above equation
for the gauge coupling gives
\eqn\invgu{
\inv{g^2} =   \inv{g_0^2}  - \inv{8 \pi^{2}} \log |u|.
}

%------------------------------------------------------------------%
% The theta angle
%------------------------------------------------------------------%
\subsec{The $\theta$-angle}

Equation \invgu\ is actually part of a holomorphic equation
\eqn\holg{
{{8\pi i}\over {g^2}} + {{\theta}\over {2\pi}}
 = 
{{8\pi i}\over {(g_0)^2}} + {{\theta_0}\over {2\pi}}
 + {{2 i}\over {\pi}}\, \log u.
}
The effect of $u$ on the $\theta$-angle of the $E_8$ Yang-Mills
can be understood as follows. 
Recall from section (2) that the imaginary part of $u$
for large $u$ is given by the phase $\int B_{56}$ of
the anti-self-dual 2-form that ``lives'' on the 5-brane.
In the M-theory context, adding an instanton on the 9-brane
affects the value of the field strength $G$ in the 10+1D bulk
since the source term \kilder\ includes a piece \HWII:
\eqn\dgd{
dG \propto \delta(x_{11}) dx_{11}\wdg \trp{F\wdg F}.
}
On the other hand, the world-volume action of a 5-brane in the bulk 
contains a term \StrOPN:
$$
\int B^{(-)}\wdg G.
$$
Taking $B^{(-)}_{56}$ ($5,6$ are the directions of the $\MT{2}$)
to be constant we get a term which is proportional
to
$$
\left(\int_{\MT{2}}B^{(-)}\right) \int d^4x\, \trp{F\wdg F}.
$$
Note that because of the $\delta$-function in \dgd\
$G$ is independent of the $x_{11}$ position of the 5-brane.
Thus, $\int B^{(-)}$ behaves like a modification of the 
Yang-Mills $\theta$-angle.

%------------------------------------------------------------------%
% finishing
%------------------------------------------------------------------%
\subsec{Back to the correlator}

As we discussed in section (3),
by holomorphy \invgu\ must be valid for all $u$.
Thus we can use the same formula for small $u$,
where the theory becomes the $E_8$ theory we are interested in.
For small $u$ the Seiberg-Witten curve becomes
$$
y^2 =x^3 - \alpha u^5
$$
which is the standard form of an $E_8$ curve.
The $E_8$ theory we are interested in is conformally invariant at $u=0$,
so it cannot have a dimensionful parameter $\alpha$.
We get rid of that by a redefinition:
$$
u = \alpha \wbu,\qquad
x = \alpha^2 \wbx,\qquad
y = \alpha^3 \wby.
$$
which gives the curve
$$
\wby^2 =\wbx^3 - \wbu^5
$$
The dimension of $\wbu$ is seen to be $6$
since $\alpha$ has dimension $(-6)$ and $u$ is dimensionless.
This agrees with the dimensional analysis in section (3).
The equation for the gauge coupling now becomes
\eqn\klu{\eqalign{
\inv{g^2} & =   \inv{g_0^2}  - \inv{8 \pi^{2}}\, \log |u| \cr 
          &= \inv{g_0^2}  + \frac{6}{8 \pi^{2}}\,
      \log\left| \frac{\alpha^{-1/6}}{\wbu^{1/6}} \right|
     \cr
}}
$\alpha^{-1/6}$ then acts as a cut-off and $\wbu^{1/6}$
is a typical mass scale of the theory.
In the notation of section (2.1) we conclude that the value of $c$
for this theory is
$$
c = - \frac{3}{4 \pi^2}.
$$
The $E_8$ vector-multiplets have a value of
$$
c = \frac{C_2(E_8)}{4 \pi^2}.
$$
Since $C_2(E_8)= 30$
we conclude that we need $\frac{30}{3}=10$ of these $E_8$ theories
to saturate the $\beta$-function.

%%% ------------------------- CUT HERE ---------------------------------%

% ===================================================================== %
% Section (5): F-theory calculation
% ===================================================================== %
\newsec{The 6D current-current correlator from F-theory}

In this section, we use the duality between F theory on elliptic
Calabi-Yau 3-folds and Heterotic String on $K3$ to compute the
effective gauge coupling of heterotic string in six dimensions.
We shall see that the result agrees completely with the
corresponding  M-theory
calculation to first order.  A second order effect which is
suppressed by a factor of the volume of the K3 and by the 
length of $\MS{1}/\BZ_2$ in calculations
in the previous section naturally emerges in the F-theory setting.
In the limit in which we extract the correlator for $V_1$, i.e. taking
the volume of K3 and the size of $\MS{1}/\BZ_2$ to infinity,
this second order effect vanishes.

We start with $V_1$ and couple it to a 6D $E_8$ SYM theory.
The gauge theory is defined with a UV cut-off, but this imposes no
problem for us since all we need is the dependence of the
IR coupling constant on the VEVs of the $V_1$ theory.
To be precise, we take the $E_8$ UV cut-off to be $\Lam$
and fix the $E_8$ coupling constant at $\Lam$.
The Coulomb branch of the $V_1$ theory has a single tensor multiplet.
We denote the VEV of its scalar component by $\phi$.
$\phi$ is the tension of the BPS string in $\MR{5,1}$.
In M-theory $\phi = x T_2$.
The mass scale of the $V_1$ theory is thus $\phi^{1/2}$.
We would like to find the dependence of the IR $E_8$ coupling constant
on $\phi$ when $\phi \ll \Lambda$. Heuristically speaking,
the running $E_8$ coupling constant will receive contributions
from loops of modes from $V_1$ of mass $\sim \phi^{1/2}$.

The set-up that we have just described arises in the heterotic string
compactified on $K3$ with a small $E_8$ instanton.
We take the $(0,23)$ embedding with a single 5-brane in the bulk
close to the 9-brane with unbroken $E_8$. The F-theory dual
has a base $B$ which is the Hirzebruch surface
$\MF{11}$ with one point blown-up  \refs{\SWSIXD,\MVII}.
$\MF{n}$ is a $\CP{1}$ bundle over $\CP{1}$. Let the area of 
the fiber $\CP{1}$ in $\MF{11}$
(i.e. the K\"ahler class integrated over the fiber)
be $k_F$ and the area of zero section $\CP{1}$ of the fibration
be $k_D$.

We blow-up a point in the zero section of the fibration of $\CP{1}$
over $\CP{1}$ (see \BerSad\ for a recent discussion).
There are 10 7-branes wrapping that zero-section and passing
through a point of the exceptional divisor. These are responsible
for the unbroken $E_8$ gauge group. Let $k_E$ be the area of the
exceptional divisor. The area of the above mentioned
7-brane locus (part with unbroken $E_8$) is $k_D$.
The K\"ahler class is
$$
k = (k_F - k_E) E + k_F D + (k_D + k_E - n k_F) F
$$
where $E,D,F$ are the cohomology classes of the exceptional divisor,
base and fiber.
\eqn\intscns{\eqalign{
E\cdot E = -1, &\qquad
E\cdot D = F\cdot D = 1,\cr
D\cdot D =  n-1, &\qquad
 E\cdot F = F\cdot F = 0.\cr
}}

A 3-brane wrapping the exceptional divisor gives a the BPS string
in $\MR{5,1}$ (corresponding to the membrane connecting
the M-theory 5-brane to the end of the world).  Its tension is
given by integrating the D3-brane tension over $E$.  Using \Polch:
$$
2\kappa^2 \tau_p^2 =2\pi(4\pi^2 \alpha')^{3-p}
$$
the tension of the BPS string is simply
$$
\phi= \pi^{1/2} k_E
$$
in the units $\kappa=1$.
The volume of the whole base is
\eqn\volbas{
V = \half k\cdot k = k_F (k_D + k_E) -\half k_E^2 -{n\over 2} k_F^2.
}
This volume is the 6D inverse gravitational constant and we have
to keep it fixed.
Although the $V_1$ modes have an effect on the gravitational constant
as well, by dimensional analysis, this effect is much smaller
than $\phi$ and behaves as $\sim \phi^2$.
How should $k_F$ depend on $\phi$, in our setting?
$k_F$ measures the tension of 3-branes wrapped on $F$.
On the heterotic side, these are elementary
strings which occupy a point on K3.
Their tension is fixed in the heterotic picture. Thus $k_F$
is independent of $\phi$.

Now we come to the gauge coupling.
To do this calculation it is convenient to imagine that $E_8$ is 
broken down to $U(8)\subset E_8$. 
The gauge kinetic term for 8 unwrapped 7-branes
of the same type is
$$
\int \tau_7 {{(2\pi \a')^2}\over {4}} \trr{\rep{8}}{F^2}\, d^8 x.
$$
We are working in the conventions
$$
\trp{T^a T^b} = \delta^{ab}, \qquad a,b=1\dots 248.
$$
For the $U(8)$ subgroup this means that
$$
\trr{\rep{8}}{T^a T^b} = {1\over {2}}\delta^{ab}.
$$
This means that for a configuration of 10 7-branes forming
an $E_8$ gauge theory the gauge kinetic term is:
$$
{1\over 8}\int (2\pi \a')^2\, \tau_7\,
\left(\sum_{a=1}^{248} F^a F^a\right)\, d^8 x.
$$
From this we read off (in units where $\kappa = 1$)
$$
{1\over {4g^2}} = {1\over 8}(2\pi \a')^2\,\tau_7
= {1\over {32}}\pi^{-3/2}.
$$
Wrapping the 7-branes on $D$ we get a 5+1D $E_8$ gauge theory with
coupling constant 
$$
{1\over {4g^2}} 
= {1\over {32}}\pi^{-3/2} k_D.
$$
 From \volbas\ we find that when $V$ and $k_F$ are kept fixed
and $k_E = \pi^{-1/2} \phi$,
the $E_8$ coupling constant is
\eqn\rung{
{1\over {g(\phi)^2}} = {1\over {8}}\pi^{-3/2}\lbr
(k_D + k_E) - k_E \rbr = {1\over {(g_0)^2}}
-{1\over {8\pi^2}} \phi.
}
We have used the fact that $(k_D + k_E)$ is fixed to first order
in $\phi$ when $V$ is fixed.  The other two terms in $V$ are higher
order corrections dual to taking $K3$ and the distance between
the ends of the world to be large in the  M-theory calculations.
Eqn\rung\ describes the running of the $E_8$ coupling constant
because  of the coupling to $V_1$.
This is in complete agreement with the result \gxt\ obtained from M-theory.

%=======================================================================%
% Discussion
%=======================================================================%
\newsec{Discussion}

We have found that for the 3+1D $E_8$ super-conformal theory with
Seiberg-Witten curve
$$
y^2 = x^3 + u^5,
$$
the 2-point $E_8$ current correlator on the Coulomb branch
satisfies:
\eqn\satis{
\ev{j_\u^a(q) j_\v^b(-q)} = \cases{
{{C_2(E_8)}\over {40\pi^2}} \delta^{ab} (q_\u q_\v -q^2 \eta_{\u\v})
   \logx{ {\Lambda\over |u|^{1/6}} }
& for $|q| \ll |u|^{1/6}$
\cr
{{C_2(E_8)}\over {40\pi^2}} \delta^{ab} (q_\u q_\v -q^2 \eta_{\u\v})
   \logx{ {\Lambda\over |q|} }
& for $|q| \gg |u|^{1/6}$
\cr
}}
where $\Lambda$ is a UV cutoff which is an artifact of Fouri\'er
transforming.

We deduced that 10 copies of the $E_8$ theory can be coupled as ``matter''
to an $\SUSY{2}$ $E_8$ SYM gauge field.

In 5+1D one can similarly find the expression
for the low-energy limit of the 5+1D correlator of
the $\SUSY{(1,0)}$ $E_8$ theory on the Coulomb branch and away from
the origin:
\eqn\curvo{
\ev{j_\u^a(q) j_\v^b(-q)} = 
-{{C_2(E_8)}\over {240\pi^2}}\,
   \delta^{ab} (q^2 \eta_{\u\v} - q_\u q_\v)
   (\Lambda^2 - \phi)\qquad
{\rm for\ } |q| \ll \phi.
}
where $\phi$ is the VEV of the scalar of the low-energy tensor multiplet.

It would be interesting to determine the correlator in the UV region
$|q| \gg |\phi|$ or, equivalently, at the fixed point $\phi = 0$.
It seems that the methods presented in this paper are not powerful
enough for that purpose. Perhaps the new developments 
\refs{\ABKSS,\WitQHB,\Lowe,\KSAB} following the M(atrix)-theory
of \BFSS, would allow one to determine this correlator.

It would also be interesting to determine the 2-point function
of the energy momentum tensor for the 3+1D and 5+1D theories.
Intriguing conjectures have been proposed in \GubKle\
for the energy momentum tensor correlator in the $\SUSY{(2,0)}$
5+1D theory. The method which was developed in \GubKle\
was to scatter gravitons off the classical black-hole solution
corresponding to the $N$ coincident 5-branes in M-theory and type-IIA.
Since the low-energy description of the degrees of freedom of $N$
5-branes is a generalization of the $\SUSY{(2,0)}$ theory 
(from $N=2$ to $N>2$) it is the hope that, at least in the large $N$
limit, the form of the correlator of the $\SUSY{(2,0)}$ theory
is reproduced by the classical solution.
It might be interesting to scatter $E_8$ gluons off the classical
CHS solution of the heterotic 5-brane \CHS\ and extract the
corresponding prediction for the current-current correlator.
%=======================================================================%
% Acknowledgments
%=======================================================================%
\bigbreak\bigskip\bigskip
\centerline{\bf Acknowledgments}\nobreak
We wish to thank Steve Gubser, Igor Klebanov, Sangmin Lee,
Sanjaye Ramgoolam and Savdeep Sethi for discussions.
The research of YKEC was supported by DOE grant DE-FG02-91ER40671.
The research of OJG was supported by a Robert H. Dicke fellowship and by
DOE grant DE-FG02-91ER40671 and the research of
MK was supported by the Danish Research Academy.

%%% ------------------------- CUT HERE ---------------------------------%

\listrefs
\bye
\end